# APPROXIMATE SOLUTIONS TO FRACTIONAL SUBDIFFUSION EQUATIONS:
## The heat-balance integral method

**JORDAN HRISTOV**

Department of Chemical Engineering

University of Chemical Technology and Metallurgy, Sofia,

8, Kliment Ohrisdsky , Blvd., Sofia 1756, Bulgaria

e-mail: jordan.hristov@mail.bg ; website: http://hristov.com/jordan

**ABSTRACT**

The work presents integral solutions of the fractional subdiffusion equation by an integral method, as an alternative approach to the solutions employing hypergeometric functions. The integral solution suggests a preliminary defined profile with unknown coefficients and the concept of penetration (boundary layer). The prescribed profile satisfies the boundary conditions imposed by the boundary layer that allows its coefficients to be expressed through its depth as unique parameter. The integral approach to the fractional subdiffusion equation suggests a replacement of the real distribution function by the approximate profile. The solution was performed with Riemann –Liouville time-fractional derivative since the integral approach avoids the definition of the initial value of the time-derivative required by the Laplace transformed equations and leading to a transition to Caputo derivatives. The method is demonstrated by solutions to two simple fractional subdiffusion equations (Dirichlet problems): 1) Time-Fractional Diffusion Equation, and 2) Time-Fractional Drift Equation, both of them having fundamental solutions expressed through the M-Write function. The solutions demonstrate some basic issues of the suggested integral approach, among them: a) Choice of the profile, b) Integration problem emerging when the distribution (profile) is replaced by a prescribed one with unknown coefficients; c) Optimization of the profile in view to minimize the average error of approximations; d) Numerical results allowing comparisons to the known solutions expressed to the M-Write function and error estimations.

**Introduction**

The subdiffusion process has been observed in many real physical systems such as highly ramified media in porous systems [1, 2, 3, 4], anomalous diffusion in fractals [5], diffusion in thick membranes [6], anomalous drug absorption and disposition processes [7], heat transfer close to equilibrium [8], etc. Subdiffusion occurs in a variety of applications as discussed [9, 10, 11, 12]. There has been a growing interest to investigate the solutions of subdiffusive equations and their properties for various reasons which include modeling of anomalous diffusive and subdiffusive systems, description of fractional random walk, unification of diffusion and wave propagation phenomenon, and simplification of the results. The common methods for solving fractional-order equations are purely mathematical [13], even tough they are approximate in nature, among them: in terms of Mittag–Leffler function [14] , similarity solutions [15] , Green's function [16,17] ,operational calculus [18] , numerical methods [19] , variational iteration method [20,21], and differential transformations [22,23] .

The present work refers to an integral solution commonly known as heat-balance integral [24] to the fractional subdiffusion equation expressed generally as

$$D_t^\mu T(x,t) = b \frac{\partial^\beta T(x,t)}{\partial x^\beta} \quad (1)$$

The solution was performed with Riemann –Liouville time-fractional derivative since the integral approach avoids the definition of the initial ($t=0$) value of the time-derivative in the Laplace transformed equation and consequent transition to Caputo derivatives. The method is demonstrated by solutions to two simple fractional subdiffusion equations (Dirichlet problems):

1. Time-Fractional Drift Equation, with $0 < \mu < 1$, $\beta = 1$ and $b = -V_\mu < 0$
2. Time-Fractional Diffusion Equation, with $0 < \mu < 1$, $\beta = 2$ and $b = a_\mu > 0$

The integral solution suggests a prescribed profile with unknown coefficients satisfying the boundary conditions at the both ends of the penetration layer. The integral approach to the fractional equation suggests replacement of the real function by an approximate profile and integration over the penetration depth. The technique was demonstrated recently [25] in a solution of a hall-time fractional equation resulting by splitting of the normal (diffusion) parabolic equation and Riemann-Liouville time-derivative and to radial subdiffusion from central point source through a sphere [26]. Both of them have fundamental solutions expressed through the M-Write function [27,28] .The solutions demonstrate some basic issues of the suggested integral approach, among them: a) Choice of the profile, b) Integration problem emerging when the distribution (profile) is replaced by a prescribed one with unknown coefficients; c) Optimization of the profile in view to minimize the average error of approximations; d) Numerical results allowing comparisons to the known solutions expressed to the M-Write function and error estimations.

**2. The integral method -mathematical formulation.**

It is assumed that the temperature (concentration) $T(x,t)$ in the semi-infinite subdiffusive material satisfies the one-dimensional fractional equation (1) with $\beta = 2$ and

$$\frac{\partial^\mu T(x,t)}{\partial t^\mu} = a_\mu \frac{\partial^2 T(x,t)}{\partial x^2} \tag{2a}$$

$$T(0,t) = T_s, \ t \geq 0 \ ; \ T(x,0) = T_\infty, \ x > 0 \ ; \ \frac{\partial T(x,0)}{\partial t} = 0, \ x > 0 \tag{2b,c,d}$$

$$d^\mu T / dt^\mu = {}_{RL}D_t^\mu T(x,t) = \frac{1}{\Gamma(1-\mu)} \frac{d}{dt} \int_0^t \frac{T(x,u)}{(t-u)^\mu} du \tag{2e}$$

As it was mentioned at the beginning, the integral approach considers a finite depth of penetration $\delta$ of where $a_\mu$ is a sort of fractional diffusion coefficient of dimensions $[a_\mu] = [m^2/\sec^\mu]$ and (2e) is the Riemann-Liouville fractional derivative of $T(x,t)$ with respect to the time $t$ [18].

The temperature (concentration) into the medium evolves in time, i.e. $\delta(t)$. Beyond the point $x = \delta(t)$, the medium is undisturbed and

$$T(x,\delta) = T_\infty, \ x > \delta \ ; \ \frac{\partial T(\delta,t)}{\partial x} = 0 \ ; \ \delta(t) = 0, \ t = 0 \tag{3a,b,c}$$

This approach is supported be experimental facts of almost sharp fronts of penetration of the diffusion substances [6, 7, 29, 30]. Moreover, the fractional diffusion equation referring sub-diffusion problems [6] the heat (mass) propagation (diffusion) is so slow [8, 30, 31] that the concept of the penetration layer becomes essential in view of the fact that it really exists [6,7,29, 30]. In accordance with the heat-balance concept at any time $t$ the integral of both sides (1a) along $\delta$ should be

$$\int_0^\delta \left[ \frac{\partial^\mu}{\partial t^\mu} T(x,t) \right] dx = a_\mu \left. \frac{\partial T}{\partial x} \right|_{x=0}^\delta \tag{4a}$$

The integral of left-side in (4a) is termed hereafter – *fractional-time Heat-balance Integral* **(FT-HBI)** [25, 26]. If the distribution $T(x,t)$ across the penetration is approximated by $T_a(x)$ depending only on $x$, $0 < x < \delta(t)$ and the boundary conditions (3a, b) and (2b) are applied, then we get a profile (distribution) expressed as a function of $x$ and coefficients depending on $\delta(t)$. Replacing $T(x,t)$ by $T_a(x,\delta)$, in (4a) we get

$$\int_0^\delta \left[ \frac{\partial^\mu}{\partial t^\mu} T_a(x,\delta(t)) \right] dx = a_\mu \left. \frac{\partial T_a(x,\delta(t))_a}{\partial x} \right|_{x=0}^\delta \tag{4b}$$

Now, the main problem is the evaluation of the fractional heat-balance integral (4b) through particular expression of $T_a(x,\delta)$ and the definition of the fractional-time derivative.

$$_{RL}D_t^\mu T_a(x,\delta(t)) = \frac{1}{\Gamma(1-\mu)} \frac{d}{dt} \int_0^t \frac{T_a(x,\delta(t),\tau)}{(t-\tau)^\mu} d\tau \tag{5}$$

The crux of the fractional heat-balance integral is the left-hand side of (4a) since the distribution $T_a(x,t)$ should satisfy the integral in average, but not the original domain equation (2a). The second principle point is the choice of the approximating profile $T_a(x)$, which is commonly expressed as a polynomial function of integer order: quadratic or cubic [24]. Here we will use a generalized parabolic profile $T_a(x,t) = b_1 + b_2(1+b_3 x)^n$ with unspecified exponent [25, 26, 32,33] and $T(0,t) = T_s$ as boundary condition at $x = 0$. Applying the boundary conditions (3a, b) and (2b) to the approximate profile we get

$$T_a(x,t) = T_\infty + (T_s - T_\infty)\left(1 - \frac{x}{\delta}\right)^n \Rightarrow \Theta_a(x,t) = \frac{T - T_\infty}{T_s - T_\infty} = \left(1 - \frac{x}{\delta}\right)^n \quad (6)$$

The approximate parabolic profile satisfies the conditions imposed at the boundaries of the penetration layer and the heat-balance integral at any value of exponent $n$ as it was discussed in [25, 26, 32, 33]. The exact value of the exponent $n$ should be defined through additional conditions addressing minimization of the approximation error over the entire domain [25,26] as it is discussed further is this article regarding the particular problems at issue.

## 3 Fractional-time Heat-balance Integral

The next important step is to evaluate the time-fractional derivative of (5) replacing $T(x,t)$ by the approximate distribution $T_a(x,t)$, that is, the following integral has to be solved

$$I_{fr} = \int_0^\delta \left[\frac{\partial^\mu}{\partial t^\mu} T_a(x,t)\right] dx \quad (7)$$

The problems was conceived and developed in [25] and the solution of (7) with $\mu = 1/2$ was expressed as $I^a_{fr(1/2)} = \frac{2}{\sqrt{\pi}} \frac{1}{n+1} \frac{d}{dt}(\delta \sqrt{t})$. Following the same technique (see details about the integration in ref. 26) at any $0 < \mu < 1$ we get from (6) and (7) that

$$I^a_{fr(\mu)} = \frac{1}{(1-\mu)} \frac{1}{n+1} \frac{d}{dt}(\delta t^{1-\mu}) \quad (8)$$

Hence, the approximation of $\partial^\mu T_a / \partial t^\mu$ by $T_a$ yields [26]

$$\frac{\partial^\mu T_a(r,t)}{\partial t^\mu} = \frac{1}{\Gamma(1-\mu)}\left[\frac{1}{(1-\mu)} \frac{1}{n+1} \frac{d}{dt}(\delta t^{1-\mu})\right] \quad (9)$$

**4. Solutions and Results**

**4.1. Penetration depth**

**4.1.1. Fractional sub-diffusion equation**

According to (4b), the boundary conditions (3b, c, d) and the fractional derivative approximation get from (7). the following equation about $\delta^s(t)$, with $\delta^s(t=0)=0$.

$$\frac{d}{dt}\left(\delta^s t^{1-\mu}\right) - a_\mu \frac{N}{\delta^s} = 0, \tag{10a}$$

$$N = n(n+1)(1-\mu)\Gamma(1-\mu) \quad ; \quad a_\mu\left(\frac{\partial \Theta_a}{\partial x}\right)_{x=0} = n/\delta^s \tag{10b,c}$$

Setting $\delta t^{1-\mu} = Z(t)$ in (18a) we get

$$\frac{d}{dt}Z^2 = 2a_\mu N t^{1-\mu} \Rightarrow Z^2 = a_\mu \frac{2N}{2-\mu}t^{2-\mu} + C_1 \tag{11a}$$

$$\delta^s = \sqrt{a_\mu}\sqrt{\frac{2N}{2-\mu}}t^{\frac{\mu}{2}} + \frac{\sqrt{C_1}}{t^{1-\mu}} \Rightarrow \delta^s = \sqrt{a_\mu}\sqrt{\frac{2N}{2-\mu}}t^{\frac{\mu}{2}} \tag{11b,c}$$

The initial condition $\delta^s(t=0)=0$ leads to $C_1=0$, which is a physically defined requirement relevant to the fact that the source providing the heat (mass) at $x=0$ is of a finite power and the heat(mass) penetrates slowly into the medium. From (11b), we have that $\delta_\mu^s \equiv \sqrt{a_\mu t^\mu}$, thus defining the new length scale of the subdiffusion.

### 4.1.2. Fractional drift equation

The integration over the penetration dept of the fractional drift equation yields

$$\int_0^\delta \left[\frac{\partial^\mu}{\partial t^\mu}\Theta_a\left(x,\delta^d(t)\right)\right]dx = \int_0^\delta -V_\mu \frac{\partial \Theta_a\left(x,\delta^d(t)\right)}{\partial x} \tag{12a}$$

With the expression (9) and after the integration of the right-hand side of (12) we get the following equation about $\delta(t)$ namely

$$\frac{d}{dt}\left(\delta^d t^{1-\mu}\right) = V_\mu N \tag{12b}$$

$$N = n(n+1)(1-\mu)\Gamma(1-\mu) \quad ; \quad V_\mu\left[(\Theta_a)_{x=\delta} - (\Theta_a)_{x=0}\right] = -V_\mu \tag{12c}$$

The solution of (12b) is

$$\delta^d(t) = c_\mu N t^\mu + \frac{C_2}{t^{1-\mu}} \Rightarrow \delta^d(t) = V_\mu N t^\mu \tag{12d,e}$$

The constant $C_2 = 0$, follows from $\delta_d(t=0)=0$ and comments exactly the same as those about (11b,c). Hence, the penetration depth scales as $\delta_\mu^d(t) \equiv V_\mu t^\mu$ and it moves with a velocity proportional to the fractional drift velocity $V_\mu \left[m/\sec^\mu\right]$.

### 4.2. Approximate parabolic profiles

With the developed expression about the penetration depths $\delta^s$ and $\delta^d$ we get the following approximate profiles of both equations at issue

- **Fractional subdiffusion equation**

$$\Theta_a^s(x,t) = \left(1 - \frac{x}{\delta^s}\right)^{n_s} = \left(1 - \frac{x}{\sqrt{a_\mu t^\mu F_n^s j_\mu^s}}\right)^{n_s}, \quad F_n^s = \sqrt{2n_s(n_s+1)}, \quad j_\mu^s = \sqrt{\frac{\Gamma(2-\mu)}{2-\mu}} \qquad (13a,b,c)$$

- **Fractional drift equation**

$$\Theta_a^d(x,t) = \left(1 - \frac{x}{\delta^d}\right)^{n_d} = \left(1 - \frac{x}{V_\mu t^\mu F_n^d j_\mu^d}\right)^{n_d}; \quad F_n^d = n(n+1); \quad j_\mu^d = \Gamma(2-\mu) \qquad (14a,b,c)$$

Both profiles can be expressed through similarity variables $\eta_s = x/\sqrt{a_\mu t^\mu}$ and $\eta_d = x/V_\mu t^\mu$ taking into account that $[a_\mu] = [m^2/\sec^\mu]$ and $[V_\mu] = [m/\sec^\mu]$

### 4.3. Calibration of the profile exponent
### 4.3.1. Fractional subdiffusion equation

The approximate should satisfy the domain equation, so the mean-square error of approximation should be minimal, namely

$$E_\mu^s(t) \equiv \int_0^{\delta(t)} \left[\frac{\partial^\mu}{\partial t^\mu}\Theta_a^s(x,t) - a_\mu \frac{\partial^2 \Theta_a^s}{\partial^2 x}\right]^2 dx \geq 0, \quad E_\mu^s(t) \to \min \qquad (15)$$

$$e_\mu^s(x,t) = \frac{\partial^\mu}{\partial t^\mu} T_a(x,t) - a_\mu \frac{\partial^2 T_a}{\partial^2 x} \quad \text{and} \quad E_\mu^s(t) \equiv \int_0^{\delta(t)} \left[e_\mu^s(x,t)\right]^2 dx \qquad (16a,b)$$

The minimization of (16b) performed in [26], with $F_n^s = \sqrt{2n_s(n_s-1)}$ and $j_\mu^s = \sqrt{(2-\mu)^{-1}\Gamma(2-\mu)}$, provided positive values $n_s > 1$ as roots. The solutions provided (irrespective of $\mu$) one trivial solution $n_1 = 0$. Within the range $0 < \mu < 0.7$: three real roots: $n_1 = 0$, $n_2 > 1$ and $n_3 < 0$, and within the range $0.7 < \mu < 0.9$, the roots are 5: $n_1 = 0$, $n_2 > 1$ and $n_3 < 0$ and two complex, $n_4$ and $n_5$. The only realistic for any $\mu$ is $n = n_2 > 1$ following the basic assumption of the integral profile. The values of the exponent and the $j_\mu$ factor as various $\mu$ are plotted in Fig.1. It is obvious, that decrease in $\mu$, augments the exponent $n_s$ and reduces the correction factor $j_\mu^s$. The latter physically implies: the penetration depth decreases as the fractional order is decreased, and *vice versa*. All calculations (from eq.3 to eq.16) were performed by Maple 13.

The nonlinear regression analysis (by Origin 6.0) yields the best correlation by the Gauss distribution, namely [26]

$$n_\mu^s \approx n_{\mu 0}^s + \frac{A}{w}\sqrt{\frac{2}{\pi}} \exp\left[\frac{(n_\mu^s - n_{\mu c}^s)^2}{w^2}\right] \Rightarrow n_\mu^s \approx 1.256 + 0.224 \exp\left[\frac{(n_\mu^s - 0.264)^2}{4.303}\right] \qquad (17a)$$

with $n_{\mu 0} = 1.256$, $n_{\mu c} = 0.264$, $w = 2.074$, $A = 0.583$; $\chi^2 = 0.00003$; $R_{res}^2 = 0.91293$ (17b)

**4.3.1.1. Fractional subdiffusion equation-Numerical experiments**

Examples of distributions generated by the approximate solution (13a,bc,) at various $0 < \mu < 1$ are shown in Fig.2a,b through the dimensionless variable $x/\sqrt{a_\mu t^\mu}$ with $a_\mu = 1$. Further, numerical results for short and large times at fixed values of the space co-ordinate $x$ are illustrated by Fig. 3 and Fig. 4, respectively. The profiles are physically adequate and show the general effect of the fractional order $\mu$ on the distance of propagation (the point where the profiles go to zero-see Fig.3a and Fig.4a) as well as the time required to attain the saturation (the plateau in the curves presented by Fig. 3b,c,d and Fig.4b,c).

**4.3.2. Fractional drift equation**

*4.3.2.1 Mean-square error approach and fixed exponents*

The condition, the fractional drift equation to satisfy the domain equation reads

$$E_\mu^d(t) \equiv \int_0^{\delta(t)} \left[\frac{\partial^\mu}{\partial t^\mu}\Theta_a^d(x,t) + V_\mu \frac{\partial \Theta_a^d}{\partial x}\right]^2 dx \geq 0, \quad E_\mu^d(t) \to \min \qquad (18a)$$

With the dimensionless profile (14a) we have (see the expression (9) for $\partial^\mu/\partial t^\mu$)

$$\frac{\partial^\mu}{\partial t^\mu}\Theta_a^d(x,t) = \frac{1}{\Gamma(1-\mu)}\left[\frac{1}{(1-\mu)}\frac{1}{n+1}\frac{d}{dt}\left(\delta t^{1-\mu}\right)\right] = V_\mu n_d \qquad (19a)$$

Because $\delta_\mu = V_\mu t^\mu F_n^d j_\mu^d$, $j_\mu^d = \Gamma(2-\mu)$ and $F_n^d = n_d(n_d + 1)$ we get

Further, the drift term is

$$\frac{\partial \Theta_a^d}{\partial x} = -V_\mu \frac{n_d}{\delta_d}\left(1 - \frac{x}{\delta_d}\right)^{n_d - 1} \qquad (19b)$$

$$e_\mu^d(x,t) = V_\mu n_d - V_\mu n_d \frac{1}{\delta_d}\left(1 - \frac{x}{\delta_d}\right)^{n_d - 1} = V_\mu n_d \left[1 - \frac{1}{\delta_d}\left(1 - \frac{x}{\delta_d}\right)^{n_d - 1}\right] \qquad (19c)$$

Hence, the error decreases as the penetration depth $\delta_d$ evolves in time. This error is always positive because

$1 - x/\delta_d \geq 0$, so the integration over the penetration depth of $\left[e_\mu^d(x,t)\right]^2$ is

$$V_\mu t^\mu j_\mu n_d (n_d + 1) + \frac{1}{V_\mu t^\mu j_\mu n_d (n_d + 1)(2n_d - 1)} - \frac{1}{n_d} \quad (20)$$

The solution to (20) with respect to $n_d$ is hard to be developed in a general form because the first two terms depends on the time $t^\mu$. The second term decreases in time, so we address the first term which becomes $2V_\mu t^\mu j_\mu$ with the minimal value of the exponent $n_d = 1$ taking into account that exponents of the parabolic profile should be $n \geq 1$. With $n_d = 1$, for example, we have a linear profile with respect to the similarity variable $\eta_d = \frac{x}{V_\mu t^\mu}$. These results are shown in Fig. 5a. Obviously, the prediction by the linear profile is unsatisfactory. Simple numerical tests with $\mu = 1/2$ (see the other curves in Fig.5a reveals that with increase in the value of $n_d$ the approximate profiles approaches the exact solution expressed through the M-Wright function [27]

$$M_{1/2} \equiv \exp-\left(x^2/4V_\mu^2 t\right), \quad (21)$$

The best fits where detected with $n_d = 1.75$ in the range of $2.25 \leq \eta_d \leq 5$ as it is discussed in the next section concerning numerical experiments. The calculations were performed with Maple 13.

Similar tests with $\mu = 1/3$ (see Fig. 5b) show that the approximate profile with $n_d = 1.25$ within the range $0 \leq \eta_d \leq 1.25$ matches the exact solution [27]

$$M_{1/3} = 3^{2/3} Ai\left(\frac{\eta_d}{3^{1/3}}\right), \quad Ai(0) = \frac{1}{3^{2/3}\Gamma(2/3)} \quad (22)$$

quite well $Ai$ is the Airy function[27]. On the other hand, where $4 \leq \eta_d \leq 5.5$ the best approximate solutions are provided by parabolic profiles with $n_d = 1.75$ and $n_d = 2$.

Therefore, these simple numerical tests indicate that the profile exponent should be of variable-order (or distributed order) depending on the similarity variable. The numerical experiments reported in the next section clearly in show the effect of the fixed exponent on the accuracy of approximation.

**4.3.2.2. Fractional drift equation: numerical tests**

In the context of the outcomes of the previous point a series of numerical test with $\mu = 1/2$ and $\mu = 1/3$ were performed to test the effect of the effect of the exponent $n_d$ on the accuracy of the approximation. The results are summarized in Table 2 and Table 3 spanning variations of the similarity variable $\eta_d$ from 0.1 to 6 and $1 \leq n_d \leq 6$. The upper limit of $n_d$ shown in Table 2 and Table 3 is about $n_d = 2.25$ and corresponds to an error of approximation $4.5\%$ over the entire range of variations in the value of $\eta_d$. Recall, the classical heat-balance integral method [24] provides solution with an average errors of $8-9\%$. The results summarized in Table 2 reveal that varying the value of the exponent $n_d$ within the range $1.75 \leq n_d \leq 2.5$ the mentioned above accuracy of approximation was attained (the values closest to the exact solution are highlighted). Increasing the value $n_d$ beyond $n_d = 2.5$ up to $n_d = 5$ reduces the error of approximation to $1.5\%$ over the range $0.1 \leq \eta_d \leq 2$ (these results are not shown in Table. 2. While the values of the optimal $n_{d(1/2)}$ decreases with increase in the value of the similarity variable $\eta_d$, the case of $\mu = 1/3$ (see Table 3) reveals that maintaining the desired accuracy of approximation the exponent $n_{d(1/3)}$ should increase as the similarity variable increases. In the sub- range $0.3 \leq \eta_d \leq 2$ the best approximation was provided by a profile with $n_{d(1/3)} \approx 1.25$ as it is illustrated by the plots in Fig.5b, too. Similar results were observed with $n_{d(1/2)} \approx 1.75$ when $\eta_d > 2.5$ and $n_{d(1/2)} \approx 1.5$ for $\eta_d < 1$. Obviously, these trial-error numerical tests reveal tendencies in the dependence $n_d(\eta_d)$ in two particular cases where the exact solutions through the M-Wright function can be easily expressed. Numerical tests with two suggested continuous function $n_d(\eta_d)$ are discussed in the next section..

*4.3.2.32.* **Fractional drift equation: numerical tests with variable-order exponents**

The idea raised by the numerical experiments with fixed exponents $n_d$ commented above led to two *ad hoc* suggested forms of the functional relationship $n_d(\eta_d, \mu)$ tested with $\mu = 1/2$ and $\mu = 1/3$, namely :

- **Hyperbolic relationship**

$$n_d(\eta_d, \mu) = n_{d0} + \frac{\mu}{\eta_d} \qquad (25)$$

- **Inverse exponential relationship**

$$n_d(\eta_d, \mu) = n_{d0} + \mu \frac{\eta_d}{\exp(\eta_d)} \qquad (26)$$

The plots of the approximate profiles with various initial values of the exponent $n_{d0}$ and the exact profiles are shown in Fig.6a-d demonstrating the more suitable approximations that those with fixed exponents. However, the problem is still open because either the initial value $n_{d0}$ or the distributed order functions are still *ad hoc* chosen ones. However, this draws a future problem beyond the scope of the present work.

Additional numerical text performed with the almost optimal fixed exponents and showing the effect of the drift velocity $V_\mu$ (Fig.7a) and the time (Fig.7b) were performed, too. All these plots are physically adequate but require further analysis of the approximation errors which is a matter beyond the scope of the present work.

**5. Final Comments**

The approximate solutions of the fractional subdiffusion and drift equations performed by the fractional-time Integral-balance method generate distributions approaching those expressed through the M-Wright function. The method is simple and easily applicable. However, the problem still open in its complete application is the choice of the exponent of the approximate parabolic profile. In the case of the subdiffusion equation the problem was solved in average by the mean-square error approach even though the cumbersome expressions of the functional relationship that has to be minimized. However, with the fractional drift equation this approached failed but there are some grounds for optimism through the distributed-order exponent. This problem is still undeveloped but some ideas could be taken from already developed solutions of the normal (Fickian) equation [32, 33] that draw future studies.

**Acknowledgments**

The work was supported by the grant N10751/2010 of Univ. Chemical Technology and Metallurgy (UCTM), Sofia, Bulgaria.

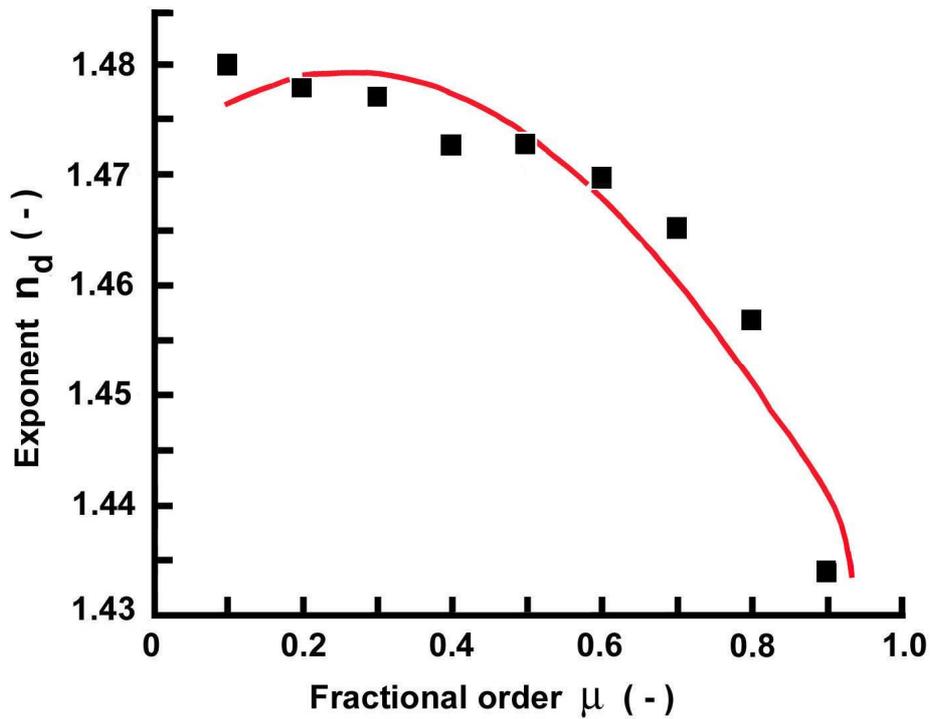

Fig.1. Optimal exponents of the parabolic profile approximating the solution of the fractional subdiffusion equation. The line corresponds to the approximating function (17a, b).

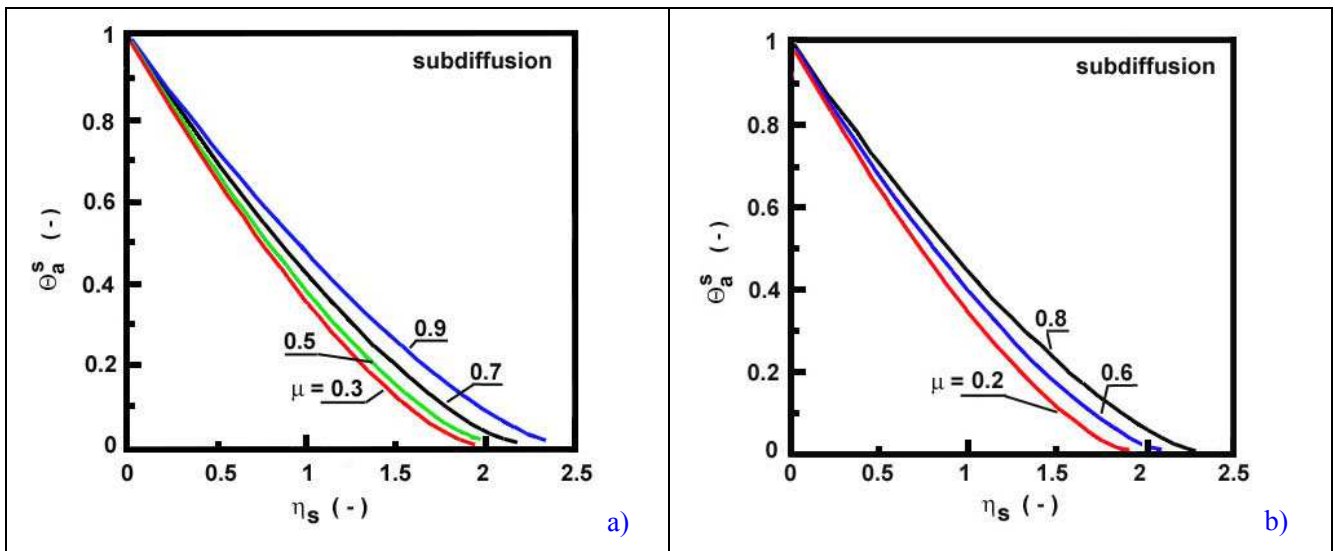

Fig.2. Approximate solutions (distributions) of the fractional subdiffusion equation generated by the profile (13a,b,c) and expressed through the similarity variable $\eta_s$. The exponents are shown in Fig. 1.
  a) Fractional orders: 0.3, 0,5, 0.7 and 0.9.
  b) Fractional orders: 0.2, 0.6 and 0.89.

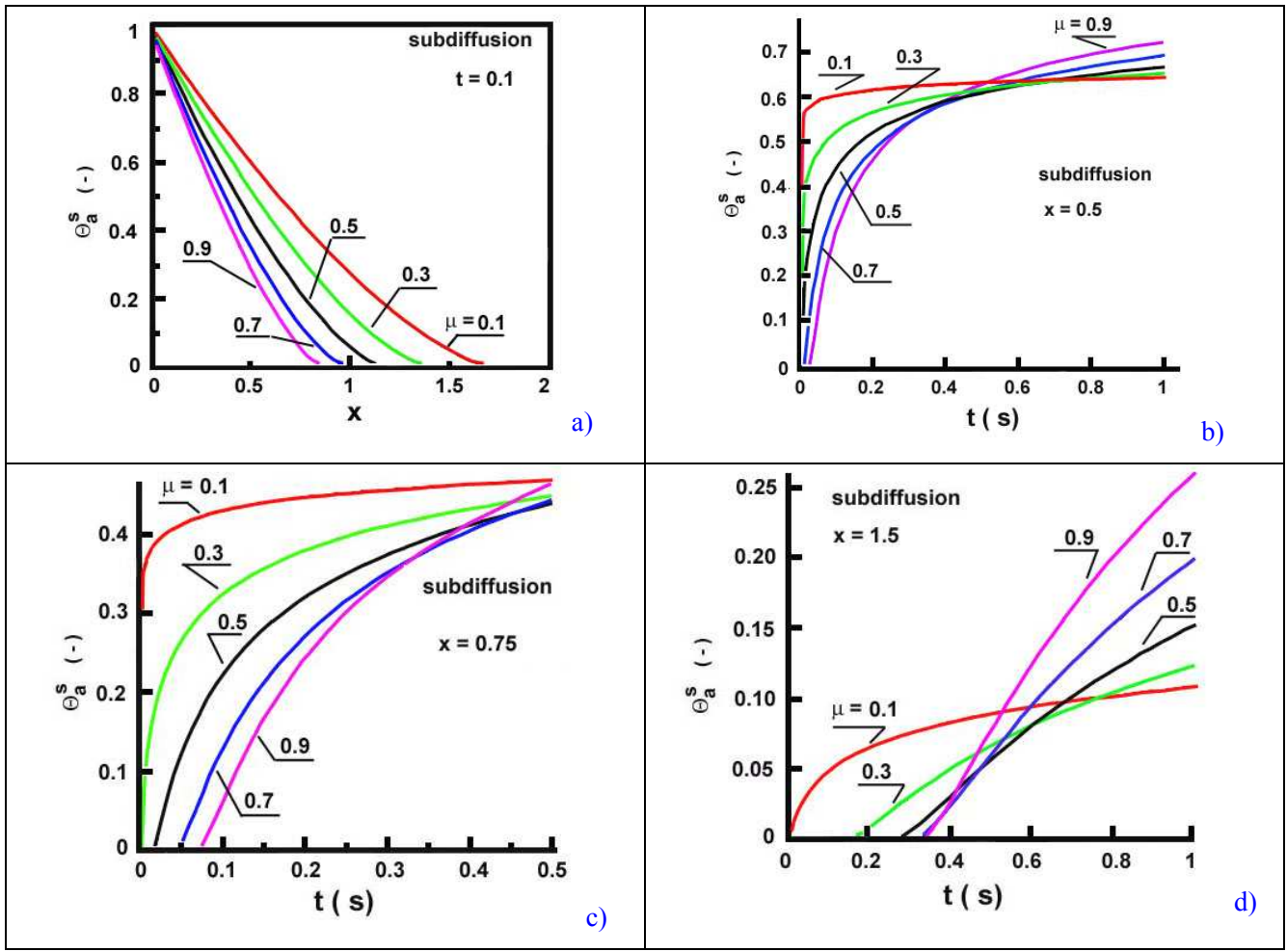

Fig.3d

Fig.3. Short-time distributions of the fractional subdiffusion equation generated by the approximate solutions. For simplicity of calculations it is assumed $a_\mu = 1$.

a) Approximate solutions as a function of the space coordinate $x$ and a fixed time $t = 0.1$
b) Time-evolution of the distributions at short distances ($x = 0.5$).
c) Time-evolution of the distributions at short distances $x = 0.75$.
d) Time-evolution of the distribution at moderate distances ($x = 1.5$).

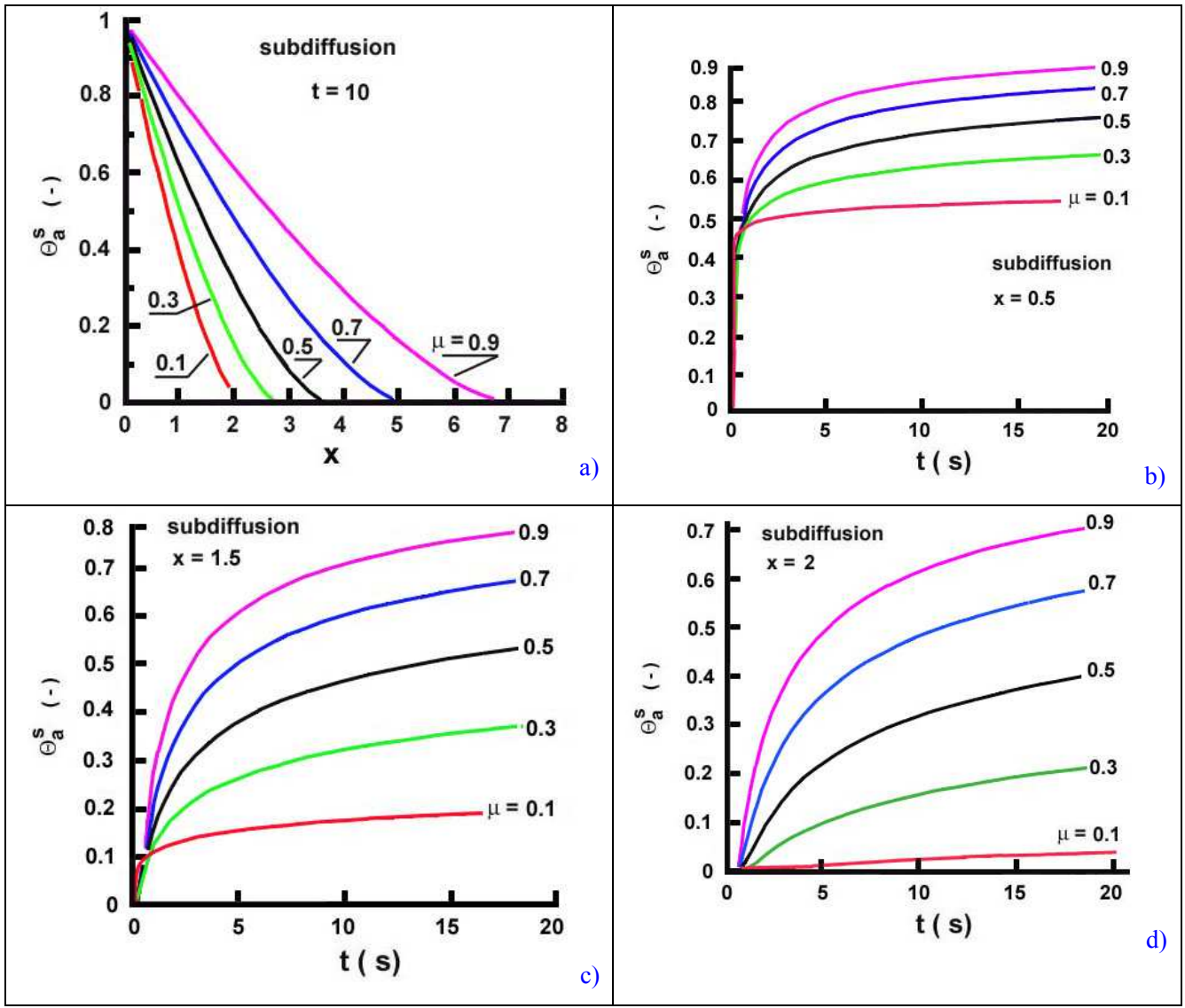

Fig.4. Large-time profiles of the distributions of the fractional subdiffusion equation generated by the approximate solutions. For simplicity of calculations it is assumed $a_\mu = 1$

a) Large-time ($t = 10$) distributions at large distances from the origin.
b) Large-time distributions at a short distance from the origin ($x = 0.5$).
c) Large-time distributions at a moderate distance from the origin ($x = 1.5$).
d) Large-time distributions at a moderate distance from the origin ($x = 2$).

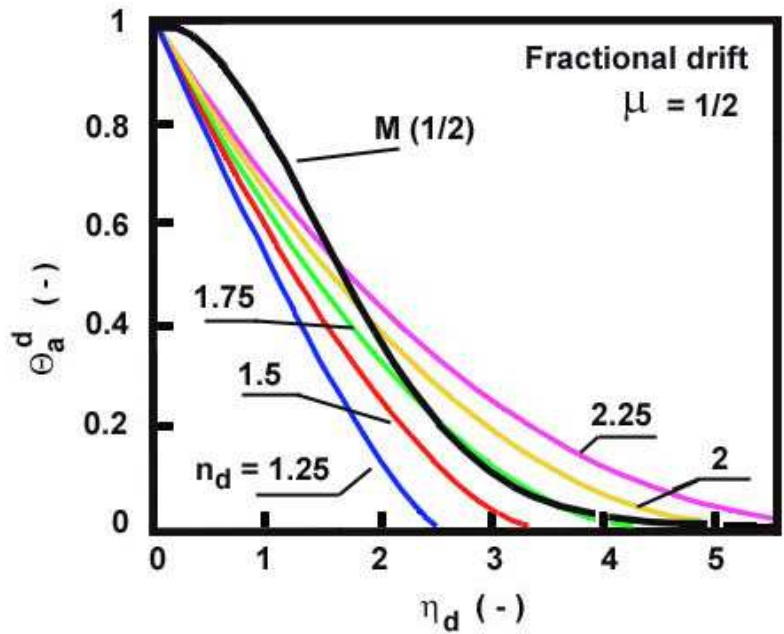

Fig.5a

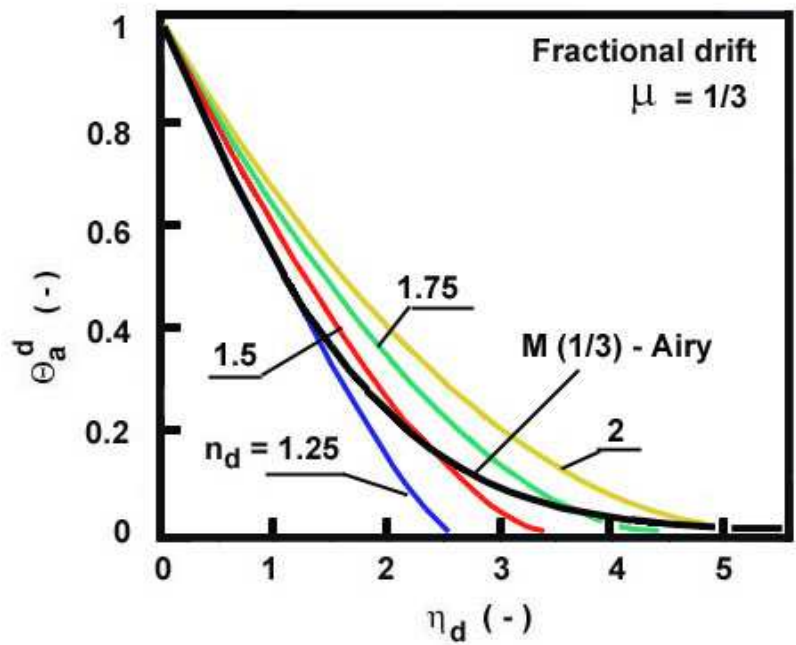

Fi.5d

Fig. 5. Approximate solutions of the fractional drift equation generated by the parabolic profile with various fixed exponents.
  a) Approximate profiles at $\mu = 1/2$.
  b) Approximate profiles at $\mu = 1/3$.

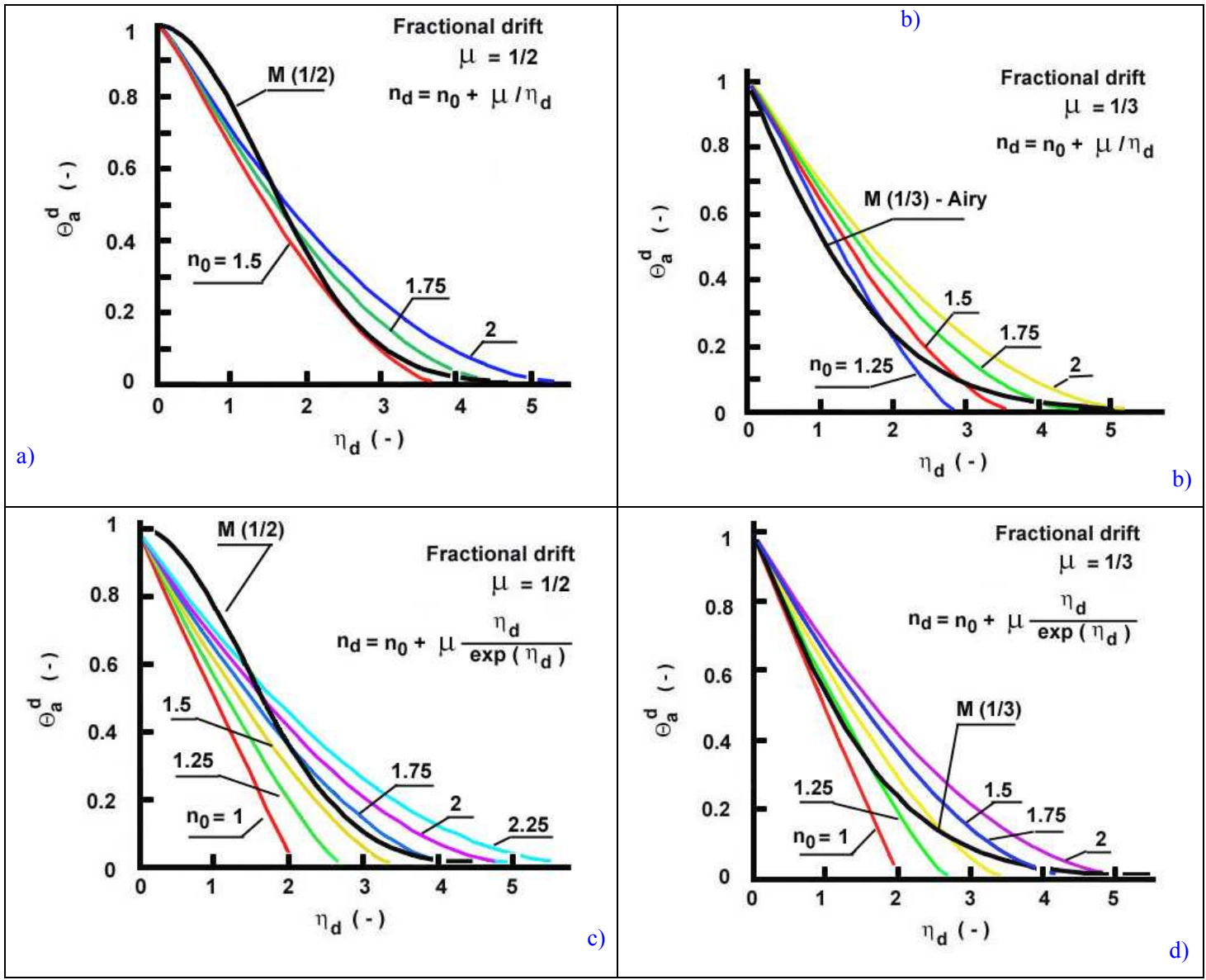

Fig.6. Approximate solutions of the fractional drift equation generated by the parabolic profile with exponents of distributed order, $n_d = f(\eta_d)$

a) Hyperbolic relationship at $\mu = 1/2$
b) Hyperbolic relationship at $\mu = 1/3$
c) Non-linear inverse exponential relationship at $\mu = 1/2$
d) Non-linear inverse exponential relationship at $\mu = 1/3$

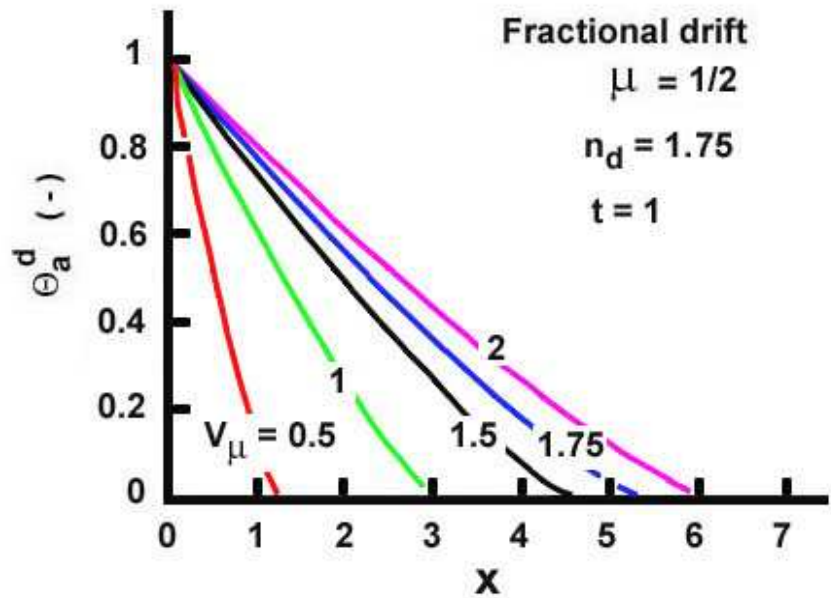

Fig.7a

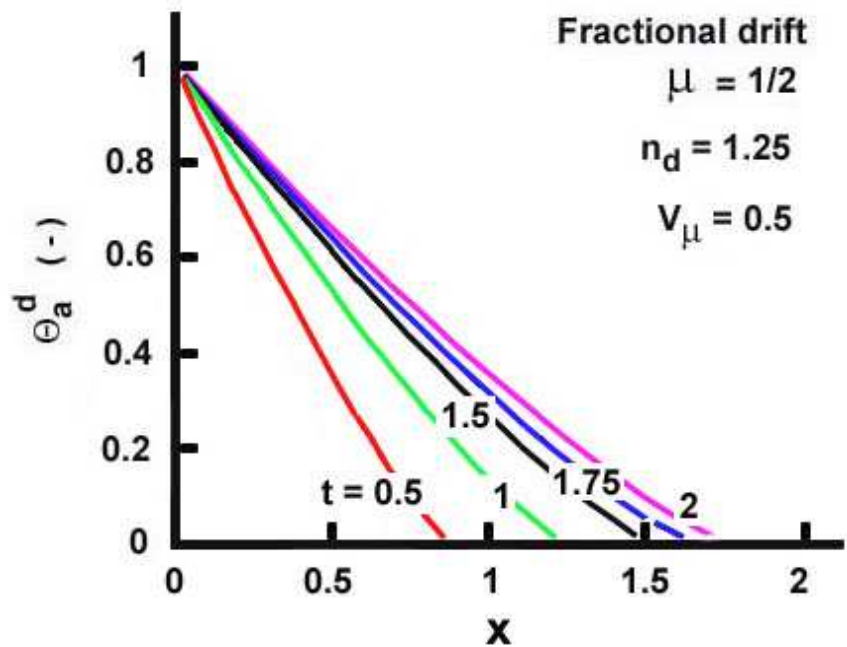

Fig.7b

Fig.7. Approximate solutions of the fractional drift equation showing the effect of the drift velocity $V_\mu$ and the time . The exponents were chosen as the best performed values- see the text about the numerical experiments.
a) Approximate profiles at $\mu = 1/2$ and short times and large distances.
b) Approximate profiles at $\mu = 1/2$ and moderate times and distances but with a low drift velocity.

Table 1. Fractional subdiffusion equation. Values of the optimal exponent (mean-square error approach) and the fractional correction factor $j_\mu$ as function of the fractional order $\mu$.

| Fractional order $\mu$ | Optimal exponent $n_s$ | fractional correction factor $j_\mu$ |
|---|---|---|
| 0.1 | 1.480 | 0.711 |
| 0.2 | 1.478 | 0.719 |
| 0.3 | 1.477 | 0.731 |
| 0.4 | 1.472 | 0.768 |
| 0.5 | 1.472 | 0.768 |
| 0.6 | 1.469 | 0.796 |
| 0.7 | 1.465 | 0.830 |
| 0.8 | 1.456 | 0.874 |
| 0.9 | 1.434 | 0.929 |

Table 2. Fractional drift equation – errors in approximation of the solutions through the heat-balance integral method. $\mu = 1/2$

| Approximate $\eta_d$ | Exact $M(1/2)$ | $n_d$ | | | | | | |
|---|---|---|---|---|---|---|---|---|
| | | 1 | 1.25 | 1.5 | 1.75 | 2 | 2.25 | 2.5 |
| 0.1 | 0.9975 | 0.9435 | 0.9501 | 0.9552 | 0.9593 | 0.9627 | 0.96561 | 0.9680 |
| 0.2 | 0.9900 | 0.8871 | 0.9007 | 0.9111 | 0.9193 | 0.9261 | 0.9318 | 0.9367 |
| 0.3 | 0.9777 | 0.8307 | 0.8518 | 0.8676 | 0.8801 | 0.8903 | 0.8988 | 0.9060 |
| 0.4 | 0.9607 | 0.7743 | 0.8035 | 0.8250 | 0.8416 | 0.8552 | 0.8664 | 0.8759 |
| 0.5 | 0.9394 | 0.7179 | 0.7558 | 0.7830 | 0.8039 | 0.8207 | 0.8347 | 0.8465 |
| 0.6 | 0.9139 | 0.6614 | 0.7087 | 0.7418 | 0.7669 | 0.7870 | 0.8036 | 0.8176 |
| 0.7 | 0.8847 | 0.6050 | 0.6622 | 0.7013 | 0.7307 | 0.7540 | 0.7732 | 0.7893 |
| 0.8 | 0.8521 | 0.5486 | 0.6164 | 0.6616 | 0.6952 | 0.7217 | 0.7434 | 0.7616 |
| 0.9 | 0.8166 | 0.4922 | 0.5712 | 0.6226 | 0.6604 | 0.6901 | 0.7143 | 0.7346 |
| 1 | 0.7788 | 0.4358 | 0.5267 | 0.5845 | 0.6265 | 0.6592 | 0.6858 | 0.7081 |
| 1.25 | 0.6766 | 0.2947 | 0.4188 | 0.4927 | 0.5449 | 0.5851 | 0.6174 | 0.6443 |
| 1.5 | 0.5697 | 0.1537 | 0.3163 | 0.4063 | 0.4683 | 0.5153 | 0.5530 | 0.5842 |
| 1.75 | 0.4650 | 0.0126 | 0.2200 | 0.3257 | 0.3968 | 0.4500 | 0.4925 | 0.5276 |
| 2.0 | 0.3678 | -0.128 | 0.1317 | 0.2512 | 0.3303 | 0.3892 | 0.4358 | 0.4743 |
| 2.25 | 0.2820 | | 0.0543 | 0.1835 | 0.2692 | 0.3327 | 0.3830 | 0.4245 |
| 2.5 | 0.2096 | | -0.004 | 0.1233 | 0.2135 | 0.2807 | 0.3339 | 0.3779 |
| 2.75 | 0.1509 | | | 0.0716 | 0.1634 | 0.2331 | 0.2886 | 0.3345 |
| 3 | 0.1053 | | | 0.0303 | 0.1192 | 0.1899 | 0.2469 | 0.2943 |
| 3.25 | 0.0713 | | | 0.0032 | 0.0810 | 0.1511 | 0.2088 | 0.2571 |
| 3.75 | 0.0297 | | | -0.046 | 0.0247 | 0.0868 | 0.1430 | 0.1916 |
| 4 | 0.0183 | | | | 0.0077 | 0.0613 | 0.1152 | 0.1631 |
| 4.25 | 0.0109 | | | | 0.00005 | 0.0402 | 0.0907 | 0.1373 |
| 4.5 | 0.0063 | | | | -0.0044 | 0.0236 | 0.0694 | 0.1141 |
| 4.75 | 0.0035 | | | | | 0.0113 | 0.0512 | 0.0934 |
| 5 | 0.0019 | | | | | 0.0035 | 0.0360 | 0.0752 |
| 5.25 | 0.0010 | | | | | 0.0001 | 0.0238 | 0.0592 |
| 5.5 | 0.0005 | | | | | -0.0011 | 0.0142 | 0.0455 |
| 6.0 | 0.00001 | | | | | | 0.0028 | 0.0243 |

Table 3. Fractional drift equation – errors in approximation of the solutions through the heat-balance integral method. $\mu = 1/3$

| Approximate | Exact | $n_d$ | | | | | |
|---|---|---|---|---|---|---|---|
| $\eta_d$ | $M(1/2)$ | 1 | 1.25 | 1.5 | 1.75 | 2 | 2.25 |
| 0.1 | 0.9494 | 0.9449 | 0.9513 | 0.9563 | 0.9603 | 0.9636 | 0.9664 |
| 0.2 | 0.8993 | 0.8899 | 0.9031 | 0.9132 | 0.9213 | 0.9279 | 0.9335 |
| 0.3 | 0.8497 | 0.8349 | 0.8554 | 0.8708 | 0.8830 | 0.8929 | 0.9012 |
| 0.4 | 0.8010 | 0.7798 | 0.8083 | 0.8291 | 0.8454 | 0.8586 | 0.8696 |
| 0.5 | 0.7533 | 0.7248 | 0.7617 | 0.7881 | 0.8085 | 0.8249 | 0.8385 |
| 0.6 | 0.7069 | 0.6698 | 0.7156 | 0.7478 | 0.7723 | 0.7920 | 0.8082 |
| 0.7 | 0.6619 | 0.6148 | 0.6702 | 0.7082 | 0.7369 | 0.7596 | 0.7784 |
| 0.8 | 0.6184 | 0.5597 | 0.6254 | 0.6693 | 0.7021 | 0.7280 | 0.7492 |
| 0.9 | 0.5766 | 0.5047 | 0.5811 | 0.6312 | 0.6681 | 0.6970 | 0.7207 |
| 1 | 0.5365 | 0.4497 | 0.5376 | 0.5938 | 0.6348 | 0.6667 | 0.6928 |
| 1.25 | 0.4442 | 0.3121 | 0.4319 | 0.5038 | 0.5547 | 0.5940 | 0.6256 |
| 1.5 | 0.3634 | 0.1745 | 0.3311 | 0.4188 | 0.4794 | 0.5254 | 0.5623 |
| 1.75 | 0.2939 | 0.0370 | 0.2361 | 0.3392 | 0.4088 | 0.4610 | 0.5026 |
| 2.0 | 0.2351 | -0.100 | 0.1484 | 0.2654 | 0.3430 | 0.4008 | 0.4467 |
| 2.25 | 0.1861 | | 0.0703 | 0.1979 | 0.2823 | 0.3449 | 0.3944 |
| 2.5 | 0.1458 | | 0.0083 | 0.1374 | 0.2267 | 0.2931 | 0.3457 |
| 2.75 | 0.1132 | | -0.028 | 0.0847 | 0.1764 | 0.2456 | 0.3005 |
| 3 | 0.0870 | | -0.079 | 0.0413 | 0.1316 | 0.2022 | 0.2589 |
| 3.25 | 0.0662 | | | 0.0099 | 0.0926 | 0.1631 | 0.2206 |
| 3.75 | 0.0374 | | | -0.031 | 0.0330 | 0.0974 | 0.1541 |
| 4 | 0.0277 | | | | 0.0134 | 0.0709 | 0.1258 |
| 4.25 | 0.0204 | | | | 0.0019 | 0.0485 | 0.1006 |
| 4.5 | 0.0149 | | | | -0.001 | 0.0304 | 0.0785 |
| 4.75 | 0.0108 | | | | | 0.0165 | 0.0594 |
| 5 | 0.0077 | | | | | 0.0068 | 0.0432 |
| 5.25 | 0.0055 | | | | | 0.0013 | 0.0298 |